\documentclass[%
aip,
amsmath,amssymb,
reprint,
]{revtex4-1}
\usepackage{chemformula} % Formula subscripts using \ch{}
\usepackage[T1]{fontenc} % Use modern font encodings
\usepackage{float}
\usepackage{subfig}
\usepackage{soul}
\usepackage{amsmath}
\usepackage{hyperref}

\usepackage{polyglossia}
\setmainlanguage{english}


\makeatletter
\def\@email#1#2{%
	\endgroup
	\patchcmd{\titleblock@produce}
	{\frontmatter@RRAPformat}
	{\frontmatter@RRAPformat{\produce@RRAP{*#1\href{mailto:#2}{#2}}}\frontmatter@RRAPformat}
	{}{}
}%
\makeatother

\begin{document}
	
	\author{David Ribar}
	\affiliation{Computational Chemistry, Lund University, P.O.Box 124, S-221 00 Lund, Sweden}
	\author{Clifford E. Woodward}
	\affiliation{School of Physical, Environmental and Mathematical Sciences University College, University of New South Wales, ADFA Canberra ACT 2600, Australia}
	\author{Jan Forsman}
	\email{jan.forsman@compchem.lu.se}
	\affiliation{Computational Chemistry, Lund University, P.O.Box 124, S-221 00 Lund, Sweden}
        %	\title{Double-layer interactions in solutions containing polyampholytes and simple salt}
	\title{Polyampholyte model of ion clusters: double-layer interactions in the presence of dissociated simple salt}        
	
	\date{\today}

%\begin{tocentry}
%\includegraphics[scale=1]{TOC.pdf}  
%\end{tocentry}
\begin{abstract}
  We explore interactions between equally charged surfaces, in the presence of simple salt and
  either neutral or monovalently charged polyampholytes. We consider the possibility of
  using these charged polymers as crude models of ion clusters. The latter have been hypothesised to
  form in concentrated aqueous salt solutions, and are possibly related to \textit{anomalous underscreening}.
  This phenomenon usually manifests itself by unexpectedly strong and long-ranged effective forces
  at very high ionic strengths. If ion clusters are formed, they are expected
  to carry at most a weak net charge. Keeping this in mind, we investigate 
  how polyampholyte chains mediate interactions between charged surfaces. A significant amount of
  simple salt is also present, in most cases.
  We highlight that if the charges of the polyampholytes are unevenly distributed, there is a
  polarisation response that in turn can generate very strong and long-ranged surface forces, even at rather high
  concentrations of simple salt. Aside from their possible relevance to ion clusters and underscreening phenomena, 
  these results also suggest the possibility of tailoring synthetic polyampholytes, in order to
  regulate colloidal stability.
    \end{abstract}
    
\maketitle

\section{Introduction}
Using simple salts to tune the stability of colloidal dispersions containing charged particles is a long-established and
theoretically well-grounded
process
\cite{Gouy:10,Chapman:13,Derjaguin:41,Verwey:48,Oosawa:71,Guldbrand:84,Kjellander:86,Nordholm:84a,Nordholm:84b,Israelachvili:91,Evans:94,Holm:01,Forsman:04c}.
For aqueous systems
with monovalent ions, mean-field theories provide reliable predictions at low and intermediate concentrations. A central
outcome of such treatments is the Debye screening length, $\lambda_D$, which characterizes the effective range of
electrostatic interactions in an ionic medium. Increasing the salt concentration enhances ionic
screening, thereby reducing $\lambda_D$ ($ \sim 1/\sqrt{c}$).

Surprisingly, recent measurements using the Surface Force Apparatus (SFA) have revealed
a non-monotonic trend: beyond a threshold concentration (typically near 1 M) the range of effective repulsion
between charged surfaces {\em increases} with added salt \cite{Gebbie:13,Gebbie:15,Smith:16,Fung:23}. Similar
trends have been reported in studies of colloidal stability \cite{Yuan:22} and thin films \cite{Gaddam:19}, though
conflicting results also exist \cite{Kumar:22}. Despite numerous theoretical efforts to rationalise these
observations
\cite{Lee:17,Coupette:18,Rotenberg:18,Kjellander:20,Coles:20,Zeman:20,Cats:21,Krucker:21,Hartel:23,Elliot:24,Forsman:24a,Ribar:25b}, the
molecular origin of this anomalous behaviour remains unresolved.

One proposed mechanism involves the formation of ionic clusters, which would alter the effective free ion
population and lead to unexpectedly long-ranged interactions at high ionic
strengths \cite{Gebbie:13,Ma:15,Hartel:23,Komori:23,Elliot:24,Ribar:24,Ribar:25b}. Experimental evidence
supporting ion clustering at high concentrations has been reported in previous
studies \cite{Sedlak:06a,Sedlak:06b, Liao2025, Irving2023, Georgalis2000, Fetisov2020, Shalit2016, Straub2022}.

In this work, we examine surface forces in aqueous environments containing (mainly) linear ionic clusters
that crudely mimic the ionic aggregates that may form in reality. We use an implicit
solvent representation, where water enters only through its dielectric
constant ($\varepsilon_r = 78.3$), and where some of the 
ions are connected to form linear aggregates. These clusters either carry a univalent net charge, or are
overall neutral. A population of simple salt (at least 100mM) is also included in the (model)
solution, keeping in mind that even if ion clusters are formed, one would
anticipate a remaining fraction of dissociated ions. These systems are
analysed using classical polymer Density Functional Theory, cDFT. We demonstrate that
such “clustered electrolytes” can generate remarkably strong repulsive forces between similarly
charged surfaces at short and intermediate separations, especially if the polyampholytes
have a heterogeneous charge distribution, allowing a strong polarisation response.

\section{Model and theory}
\begin{figure*}
	\centering
	\includegraphics{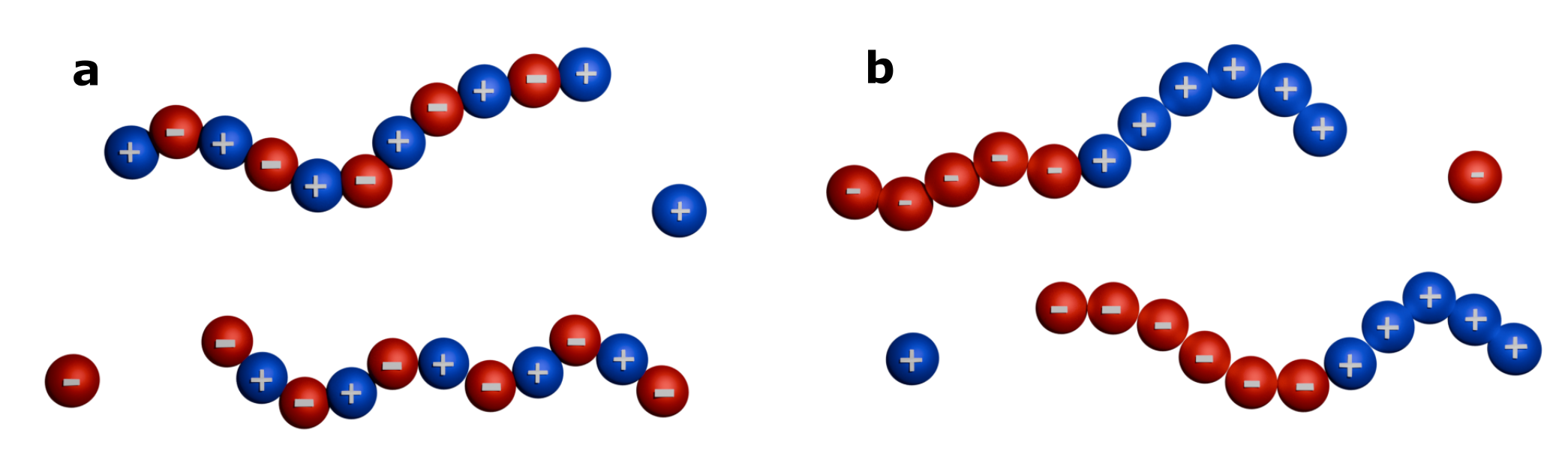}
	\caption{An illustration of a (monovalent) 11:11 polyampholyte plus 1:1 simple salt. \textbf{a} The {\em alternating} charge model.
					\textbf{b} The  {\em block} charge model.}
	\label{fig:cartoon}
\end{figure*}

%Let us first agree on the model, and a nomenclature.
The cartoon in Figure \ref{fig:cartoon} illustrates our two main models, which 
contain (monodisperse) ion clusters plus simple salt.

Figure \ref{fig:cartoon} shows monovalent linear 11-mer polyampholytes, which we
denote as a 11:11 salt, keeping in mind that this does {\em not}
imply that the species are multivalent. Indeed, we will restrict our models to contain only monovalent or
net neutral polyampholytes.  The single charged monomers are
connected by bonds with a fixed length, $b$, but with full rotational freedom.
While we primarily use linear architectures for these polyampholytes, some
comparisons with results from branched chain models are made in the Appendix.
We find that, at least for alternating charge models, cDFT
predictions with linear and branched chains agree almost quantitatively. 

We shall consider polyampholytes with two different charge arrangements: {\em alternating} charges and
{\em block} charges. Even degrees of polymerisation give
neutral polyampholytes, whereas odd degrees of polymerisation leads to
a monovalent polyampholyte salt. We use $r$ to denote the degree of polymerisation (the number of
monomers per polyampholyte chain). It should be noted that in a ``real'' solution, the
{\em block} charge model might be prone to folding and flocculation at low concentrations
of simple salt. However, these polymers are likely to be soluble at high ionic strengths.

Unless otherwise specifically stated, surface interactions will be evaluated for like-charged walls with a surface charge density
of $\sigma_s =  -0.014286 e/${\AA}$^2$. This is close to estimated values for mica in salt solution \cite{Crothers:21}, which is used in SFA experiments. 
%, and the separation is denoted by $h$.
The calculations are performed using polymer-cDFT, which accounts for  hard-sphere exclusion between particles 
and with a mean-field treatment of Coulomb interactions.

Non-bonded interactions between monomers are described by the pair potential $\phi_{ij}(r)$, defined as

\begin{equation}
		\beta\phi_{ij}(r) = \left\{
	\begin{array}{ll}
		\infty ; & r \leq d \\
		l_B \frac{Z_i Z_j}{r}; & r > d  \\
	\end{array}
	\right. ,
	\label{eq:coulomb}
\end{equation}
where $\beta = (k_BT)^{-1}$ is the inverse thermal energy, and $d$ denotes the monomer diameter. The Bjerrum length is $l_B \approx 7.16$~{\AA}, and
$Z_i$ and $Z_j$ denote the valencies of monomers $i$ and $j$ ($|Z| = 1$ throughout this work).
We use $n_+$ and $n_-$ to denote densities of positive and negative polyampholyte monomers, whereas
$n_c$ and $n_a$ are the concentrations of dissociated cations and anions.

We consider a system consisting of two infinite, rigid, and planar surfaces separated by a distance $h$.
Each surface carries a uniform charge density and the 
surfaces are immersed in the model mixture, and the inter-surface distance $h$ is varied.
The system is described within the grand canonical ensemble, implying that the confined fluid remains in chemical equilibrium
with a bulk reservoir of infinite extent. For each separation, the equilibrium (minimal) free energy is determined,
and the resulting osmotic pressure is evaluated either by numerically differentiating the free energy or
via the contact densities of monomers at the surfaces.
All polymer configurations are included following Boltzmann weighting, assuming that density variations occur solely in
the direction perpendicular to the walls \cite{Woodward:1991}.

In a completely ideal bulk polymer solution, where the only constraints are the intramolecular bonds and
no intermolecular interactions or external potentials are present, the free energy ${\cal F}_p^{id}$
for an $r$-mer can be written exactly as
\begin{equation}
\beta {\cal F}^{(id)} = \int N({\bf R}) \left(\ln [N({\bf R})] - 1 \right) d{\bf R} + \beta \int N({\bf R}) V_b({\bf R}) d{\bf R} ,
\label{eq:idpol}
\end{equation}
where ${\bf R} = ({\bf r}_1,...,{\bf r}_{r})$ denotes a polymer configuration, and $N({\bf R})$
is the distribution function such that $N({\bf R})d{\bf R}$ gives the number of polymer molecules
with configurations in the range $[{\bf R}, {\bf R}+d{\bf R}]$.
The bond potential between adjacent monomers is denoted $V_b({\bf R})$.
In this work, we assume bonds of fixed length, i.e.
\begin{equation}
	e^{-\beta V_b({\bf R})} \propto \prod \delta(|{\bf r}_{i+1}-{\bf r}_i|-b),
\end{equation}
where we recall that $b$ is the bond length, and $\delta(x)$ is the Dirac delta function.
For the simple 1:1 salt, the corresponding ideal free energy, $F_s$, naturally lacks the polymer components:
\begin{equation}
  \beta F_s^{id} = \int n_c({\bf r}) \left(\ln [n_c({\bf r})] - 1 \right) d{\bf r} + \int n_a({\bf r}) \left(\ln [n_a({\bf r})] - 1 \right) d{\bf r}.
\end{equation}
%where $n_c({\bf r})$ and $n_a({\bf r})$ denote the (unconnected) cation and anion densities.

In order to avoid complicated indexing, we will limit our cDFT description to
models of net monovalent polyampholyte chains (plus simple salt). The simplification that
ensue for net neutral polyampholytes should be obvious. 

The Helmholtz free energy for the bulk polyampholyte salt model can be expressed as
\begin{equation}
	\begin{split}
{\cal F}(V,T,N_+,N_-) =& \sum_{i=\pm}{\cal F}^{(id)}+ F_s^{id} \\
&+{\cal F}_{HS}^{(ex)}(n_+,n_-,n_c,n_a) + {\cal U}.
	\end{split}
\label{eq:grandpot}
\end{equation}
Here, $i$ distinguishes between two oppositely charged polymer species with {\em alternating} monomer charges.
Cationic chains start and end with positive monomers, whereas anionic chains terminate with negative ones.
In the bulk, both species contribute identically to the ideal free energy, while in confined or heterogeneous environments
their free energies differ. Excluded-volume effects are represented by ${\cal F}_{HS}^{(ex)}$,
which depends on the total cationic and anionic monomer densities, $n_+$ and $n_-$, as well as the densities of the
unconnected (simple salt) ions, $n_c$ and $n_a$.
This contribution is estimated using the Generalised Flory–Dimer (GFD) approach
\cite{Wichert:96}, which accounts for differences in excluded volume between
end monomers, dissociated monomers (simple salt ions), and inner monomers.
All electrostatic terms are contained in ${\cal U}$ and are treated within the mean-field approximation.

In the slit geometry, we employ the grand potential $\Omega$, related to ${\cal F}$ by the Legendre transformation
\begin{widetext}
\begin{equation}
  %\begin{eqnarray}
  \begin{split}
\Omega(V,T,\mu_+,\mu_-;[n_+,n_-,n_c, n_a\sigma])A^{-1}  = & \quad  {\cal F}(V,T;[n_+,n_-,n_c,n_a,\sigma,\Psi_D])A^{-1} \\ 
& + \sum_{i=\pm}\int (V_{ex}(z)-\mu_i) N_i(z_1..,,z_r) dz_1,..,dz_r \\
& + \int (V_{ex}(z)-\mu_c) n_c(z) dz+
\int (V_{ex}(z)-\mu_a) n_a(z) dz,
\end{split}
\end{equation}
\end{widetext}
%\end{eqnarray}  
where $z$ denotes the coordinate perpendicular to the surfaces, $A$ is the surface area, and $\Psi_D$ is the Donnan potential
that ensures overall electroneutrality.
The total free energy ${\cal F}$ encompasses contributions from configurational entropy, excluded-volume interactions,
and mean-field electrostatics, including the wall–wall interaction.
The chemical potentials of cationic and anionic chains are represented by $\mu_+$ and $\mu_-$, respectively, and the
notation for simple ions is similar: $\mu_c$ and $\mu_a$.
The surface interaction potential $V_{ex}(z)$ accounts for non-electrostatic exclusion and is purely steric,
being zero within the accessible region ($d/2<z<(h-d/2)$) and infinite elsewhere.
The equilibrium grand potential, $\Omega_{eq}(h)$, is obtained by numerical minimisation using
Picard iteration. Defining $g_s(h) \equiv \Omega_{eq}/A - p_b h$, where $p_b$ is the bulk pressure,
the net interaction free energy is given by $\Delta g_s \equiv g_s(h) - g_s(h \to \infty)$.
The tail of this interaction is assumed to be an exponential
decay (which is confirmed numerically below). We then set $g_s(h \to \infty)\approx g_s(h_0)$, where $h_0$ is a surface separation
larger than 15 decay lengths (as estimated from the fit). 

Further details of the numerical algorithm are available in \cite{Forsman:07b,Nordholm:18}, but a short summary is provided here.
Due to planar symmetry and the mean-field approximation, densities and interactions can be integrated along
the $(x,y)$ directions parallel to the surfaces while maintaining electroneutrality.
Accordingly, only $z$-dependent quantities are retained within the slit region.
The laterally integrated Coulomb interaction between two monomers with valencies $Z_\gamma$ and $Z_\delta$
at positions $z$ and $z'$ is given by $\beta\phi(z,z') = -2\pi Z_\gamma Z_\delta l_B |z-z'|$.
We introduce $\lambda_{hs} \equiv \partial{\cal F}_{HS}^{(ex)}/\partial{n_+} = \partial{\cal F}_{HS}^{(ex)}/\partial{n_-}$,
noting that cationic and anionic monomers are of equal size.

Polymer configurations are treated within a classical density functional framework under mean-field weighting.
The equilibrium monomer density is expressed in terms of auxiliary propagator functions $c(i,z)$,
where $i$ labels the monomer along the chain. Separate propagators, $c_+(i,z)$ and $c_-(i,z)$,
are introduced for cationic and anionic chains, respectively.
For cationic polymers, the contribution to the total monomer density $n'(z)$ at position $z$ reads
\begin{equation}
n'(z) = e^{\beta \mu_+} \sum_{i=1}^r c_+(i,z)*c_+(r+1-i,z),
\end{equation}
where the prime indicates that only positively charged chains are considered.
Propagator relationships will of course depend on the polymer architecture. For instance, if
the chains have alternating charges, then the recursive relation governing the auxiliary functions is
\begin{widetext}
\begin{equation}
c_+(i,z) = e^{\beta\lambda_{hs}(z)/2}\rho_+(z)^{(-1)^{(i-1)}}\int c_+(i-1,z')T(z,z')\rho_+(z')^{(-1)^{i}}dz',
\end{equation}
\end{widetext}
for $i>1$, and
\begin{equation}
c_+(1,z) = e^{\beta\lambda_{hs}(z)/2}\rho_+(z),
\end{equation}
where $\rho_+(z)\equiv e^{-\beta e\Psi(z)/2}$ and $\Psi(z)$ is the local electrostatic potential
(including $\Psi_D$).
The kernel $T(z,z')$ imposes the fixed bond length constraint and includes a normalisation factor:
\begin{equation}
T(z,z') = \frac{\Theta(|z-z'|-b)}{2b}.
\end{equation}
Anionic chains are treated analogously, and for unconnected ions there is no need for
auxiliary functions, as there are no bond integrals\cite{Nordholm:18}. 
To ensure stable convergence, conservative Picard iterations are applied—mixing a small fraction of updated densities
with the previous ones—and the Donnan potential is simultaneously adjusted to maintain global electroneutrality.
Convergence is achieved when the maximum relative change between successive iterations falls below $10^{-7}$ at all points.

In order to facilitate direct comparisons with SFA/AFM measurements, we will
assume the validity of the Derjaguin Approximation, and present interaction free
energies as force per radius, $F/R$.

\section {Results. 1. Neutral polyampholytes}
We will initially explore surface interactions in the presence of net neutral polyampholytes, i.e.
those composed of an even number of charged monomers. 

\subsection{Ideal dimers (zwitterions)}
We begin by establishing the dipolar response of a fluid of  ion pairs with opposite charges
bonded at a distance $b$  without hard cores. We denote these as ``ideal dimers''.

\subsubsection{Dielectric response, no salt}
We can evaluate the impact that these ideal dimers have
on the overall dielectric response of
the system \footnote{Recall that the implicit solvent imposes a uniform dielectric constant $\varepsilon_w$.}
by calculating the net pressure between oppositely charged surfaces. 
In order to avoid saturation effects, that are likely to vanish in the presence of added salt anyway, we
will in this case model these surfaces as weakly charged, $|\sigma_s| =  0.001 e/$ {\AA}$^2$.
Since there is no salt present, these oppositely charged walls
will experience a constant attraction at long range, with a net pressure that is inversely proportional to the
dielectric constant of the medium that separates them.
\begin{figure*}
	\centering
	\includegraphics[scale=1]{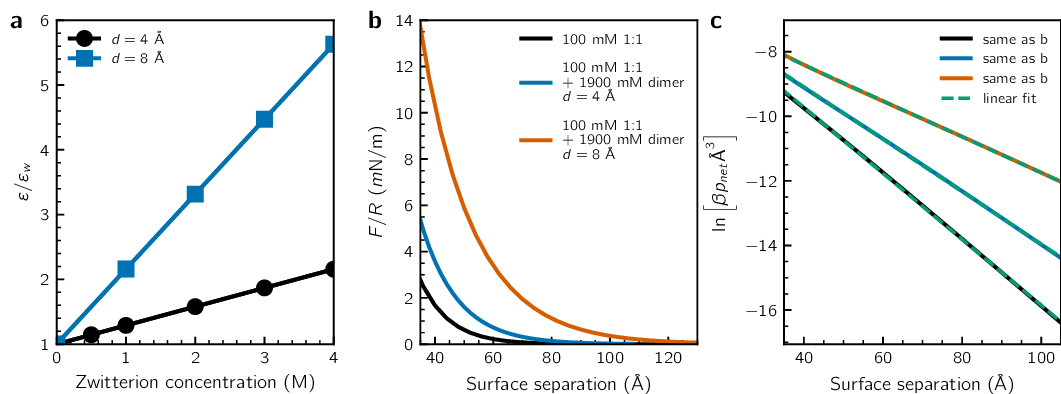}
	\caption{\textbf{a} The dependence of $\varepsilon/\varepsilon_w$ with
		      zwitterion concentration, for two different choices of $b$.
	      	  Calculated data are indicated by symbols. The lines are
		      linear regression fits.
		      \textbf{b} Force per radius values between equally charged surfaces, in the presence of
		      1900 mM zwitterions and 100 mM simple 1:1 salt (no excluded volume). \textbf{c} The
                      corresponding logarithm of the net pressure.
		      }
	\label{fig:dimer_response}
\end{figure*}
In the presence of zwitterions, this dielectric constant will shift from the reference
value $\varepsilon_w$, in the absence of dipoles, to some higher value: $\varepsilon$.
Hence, the ratio between these attractive pressures will directly give us the
ratio $\varepsilon/\varepsilon_w$. The results, for various zwitterion concentrations, and
charge separation values ($b$), are collected in Figure \ref{fig:dimer_response}\textbf{a}.
In agreement with experimental findings\cite{Edsall:35,Govrin:18}, there is a linear dependence on
concentration.

\subsubsection{Ideal dimers and ideal salt}
We now consider interactions between equally charged
surfaces with the {\em standard} charge density of $\sigma_s =  -0.014286  e/${\AA}$^2$), in the presence of
1900 mM ideal dimers and 100 mM of a simple 1:1 (ideal) salt. These concentrations are
commensurate with a salt solution which is overall 2M but where a substantial
degree of ion pairing occurs, leaving only 100 mM of dissociated salt.
We note that, in this case, no charged species will carry a hard core.
In Figure \ref{fig:dimer_response}\textbf{b}, we see the effect of the zwitterions on the double layer repulsion,
which is stronger when the zwitterions are present.  This response is also affected by 
the bond-length of the dimers.
From mean-field theory, we expect that the Debye length, $\lambda_D$, will
scale as the square root of the dielectric constant, $\varepsilon$.
Extracting the appropriate values from Figure \ref{fig:dimer_response}\textbf{a} we
find $\varepsilon(b =4$ {\AA}$,1900 mM)/\varepsilon_w\approx1.55$ and $\varepsilon(b=8$ {\AA}$,1900 mM)/\varepsilon_w\approx3.20$. 
This implies that:
\begin{equation}
  \frac{\lambda_D\text{(b=4\,{\AA},1900 mM)}}{\lambda_D\text{(no dipoles)}} \approx \sqrt{1.55}  \approx 1.25  
\end{equation}
and:
\begin{equation}
  \frac{\lambda_D\text{(b=8\,{\AA},1900 mM)}}{\lambda_D\text{(no dipoles)}} \approx \sqrt{3.20}  \approx 1.79.
\end{equation}
These ratios can be directly compared with the observed
electrostatic decay lengths, evaluated from inverted slopes of
the logarithm of the net pressure, i.e. the dashed lines in Figure\ref{fig:dimer_response}\textbf{c}.
Denoting the inverse of these linearly fitted slopes by
$\alpha$, we obtain
\begin{equation}
  \frac{\alpha\text{(no dipoles)}}{\alpha\text{(b=4\,{\AA},1900 mM)}}\approx1.26
\end{equation}
and
\begin{equation}
  \frac{\alpha\text{(no dipoles)}}{\alpha\text{(b=8\,{\AA},1900 mM)}}\approx1.84.
\end{equation}
Given the excellent agreement between these
sets of values, we conclude that the increased double layer repulsion observed in the presence of
the dimers can be fully ascribed to their dipolar response.
This response, which obviously is quite strong, is often
neglected in theoretical considerations of ion pairing, and its impact on surface forces.

\subsection{Neutral polyampholytes plus simple salt. The ``gradual growth'' of neutral ion clusters.}
As stated earlier, the above solution can
be viewed as a crude model wherein a 2 M 1:1 salt solution
undergoes partial ion pairing, leaving 100 mM of dissociated salt.
Let us now explore what happens if the clusters that are formed
are larger. Specifically, we will assume that $r$-mers are now formed
from the 2 M 1:1 solution, with 100 mM free salt remaining. For instance, with
$r=10$, we will have a mixture of 100 mM 1:1 salt and 380 mM of
neutral 10-mers. Furthermore, we will assume all species have hard cores 
of diameter $d=4$ {\AA} in our ongoing calculations. Connected monomers 
in the $r$-mers are separated by a bond-length of $b=6$ {\AA}.

\subsubsection{Chains with {\em alternating} charges}
\begin{figure*}
	\centering
	\includegraphics[scale=1]{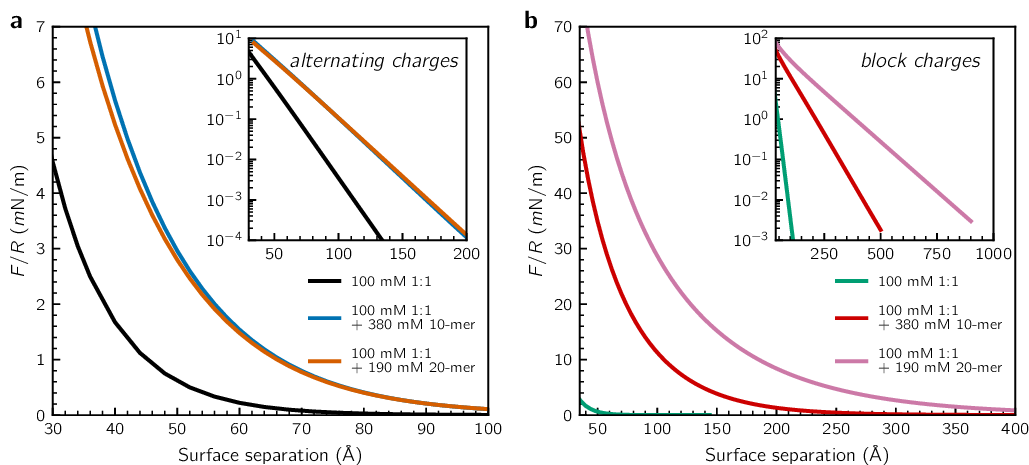}
	\caption{Interactions between equally charged surfaces, in the presence of
	     neutral $r$-mer polyampholytes, and 100 mM simple 1:1 salt (including excluded volume).
		     For a given polymer length $r$, the polyampholyte concentration is chosen in a way that is commensurate
		      with clustering of a 2 M simple salt solution.
		      The case of a pure 100 mM simple salt solution is shown as reference. The insert presents
                      the logarithm of the $F/R$ data.
		      \textbf{a} The chains are composed of monomers with an {\em alternating} charge structure.
		      \textbf{b} The chains are composed of monomers with a {\em block} charge structure.}
	\label{fig:neutral_poly}
\end{figure*}
In Figure \ref{fig:neutral_poly}\textbf{a} we
show the surface interaction free energies for systems with neutral $r$-mers containing {\em alternating} charges.
As in the earlier case for ion pairs and simple salt, the presence of the
neutral clusters leads to a significantly stronger repulsion when compared with the
simple salt only solution.  This is due to the enhanced dielectric response of the 
ionic chains.  On the other hand, an increase of $r$ from 10 to 20, even with a concomitant drop of the
polymer concentration, has essentially no impact on the net surface interaction.  This indicates the
polarization response of the chains depends primarily on the number of bonded ions, rather than the 
concentration of the clusters.

\subsubsection{Chains with {\em block} charge structure}
The corresponding contribution from polyampholytes that
are composed of two {\em blocks} of opposing charge is {\em enormous}.
This is illustrated in Figure \ref{fig:neutral_poly}\textbf{b}.
Such a complete charge separation is arguably an upper limit in ``real'' clusters,
although one would expect a significant polarisation response in the
vicinity of the charged surfaces (a more realistic model, to be
explored in future work)  At this stage, we note that the double layer forces
acting in these systems are quite sensitive to the
internal structure of the polyampholytes, i.e. our
model of ion clusters. In contrast to the case with {\em alternating}
charges, we here see a strong dependence on polymer length, $r$.
An increased degree of polymerisation leads to a stronger {\em and} more
long-ranged interaction when the chains have a {\em block} structure.
This is not surprising, since the {\em block} architecture creates 
larger dipole moments, due to an increased separation
of larger charged blocks within each cluster.  Both the separation and and size of 
the charged blocks will increase approximately with the inverse of concentration.
This means the dielectric response will also increase with the inverse concentration.    

We have already mentioned that in a bulk solution, spontaneously formed ion clusters
are unlikely to resemble {\em block} chains. Nevertheless, near
a strongly charged surface, this is perhaps not a totally unrealistic description. Moreover, these
predictions, of surface forces with a truly remarkable range and strength, even
with short polymer lengths, might imply novel applications of {\em synthesized} polyampholytes
with a {\em block} charge structure. 

Another property that might be relevant, but is
neglected in the current work, is polydispersity effects.
Real solutions of clustering ions will contain
clusters of varying size, some of which are likely to be neutral, whereas
others are weakly charged.
Polydispersity effects are beyond the scope of the current
study, but will be investigated in future work.

\section{Results. 2. Monovalent polyampholytes plus simple salt}
In a recent work\cite{Ribar:25a}, we demonstrated that monovalent polyampholyte salts
generate {\em much} stronger repulsive surface forces, than a simple 1:1 salt
at the same salt concentration. We have also established (above) how the presence
of neutral ``ion clusters'', as modelled by our polyampholyte description, also
leads to a substantially increased repulsion. Here, we will explore the surface force
response to the presence of monovalent charged polyampholytes in addition to simple salt.
We note that earlier simulation work has established that 
ion clusters tend to be either neutral, or weakly charged \cite{Hartel:23, Ribar:25b}. This is
also expended on physical grounds, since the formation of highly charged clusters
would be associated with a considerable electrostatic free energy cost.

\begin{figure*}
	\centering
	\includegraphics[scale = 1]{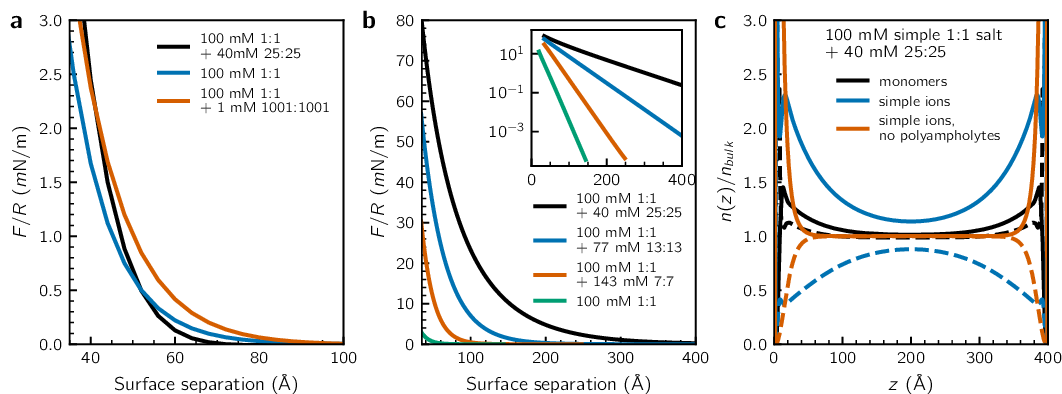}
	\caption{\textbf{a} Surface forces, in the presence of 100 mM pure 1:1 salt solutions, as well as
		      in mixtures of 100 mM simple 1:1 plus added monovalent 25:25 (40 mM) or 1001:1001 (1 mM) salt.
		      The concentrations of the polyampholyte salt are commensurate with an overall
		      monomer concentration of 2000 mM.The {\em alternating} charge
		      model has been adopted for the polyampholyte chains. \textbf{b} Same as \textbf{a}, but for 40 mM
                      monovalent 25:25 salt using the {\em block} charge structure. \textbf{c} Density
                      profiles, at $h=400$ {\AA}, in solutions containing 100 mM simple 1:1 salt and 40 mM 25:25. Full
                      lines denote the cations, while the dashed lines denote the anions.}
	\label{fig:monovalent_poly}
\end{figure*}
In Figure \ref{fig:monovalent_poly}\textbf{a}, we see how the response to the addition of
monovalent polyampholyte chains, with and {\em alternating} charge is considerably weaker
than when these are net neutral (cf. Figure \ref{fig:neutral_poly}\textbf{a}). Longer chains, at a constant monomer
concentration, do increase the range somewhat, but the net increase of the repulsion
is hardly dramatic.

On the other hand, when the polyampholytes have a {\em block} charge structure, they
generate an {\em enormous} repulsion, even at high ionic strength
and with relatively short chains. 
This is illustrated in Figure \ref{fig:monovalent_poly}\textbf{b}, where we also note that the range {\em and} the strength of the surface interaction
increase quite substantially, when such polymers are added. As we have seen earlier, this is in turn
related a strong polarisation of molecules with such an architecture. 

The slow decay of the electrostatic potential, obtained in the presence of monovalent {\em block} charge polyampholytes,
can also be illustrated by density profiles of charged species. In Figure \ref{fig:monovalent_poly}\textbf{c}, we see
how the polymer presence generate a remarkable extension to the thickness of the ``diffuse'' layer, i.e. the
electrostatic potential decays much slower than in a corresponding salt solution without polyampholytes.
A seemingly peculiar behaviour is that the anionic monomers approaches the
mid plane value from {\em above}, even though the surfaces carry a negative charge. This is not due to
overcharging, as is clear from the fact that the dissociated anions displays an opposite trend.
The behaviour should rather be considered a consequence of connectivity, since the cationic
part of the polymers are strongly adsorbed at the walls.

\subsection{Modelling the ``gradual growth'' of ion clusters with a {\em block} charge structure}
\begin{figure}
  \centering
	\includegraphics[scale = 1]{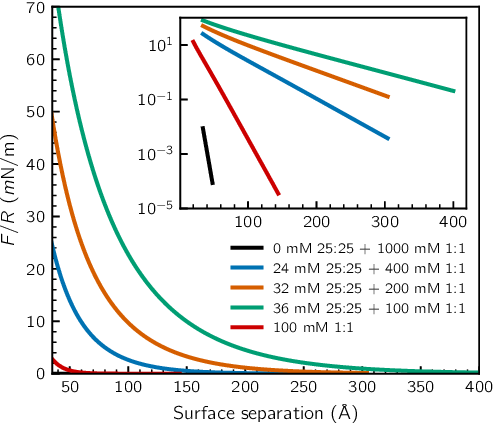}
  \caption{Surface forces, in the presence of pure 1:1 salt solutions, and
      in mixtures of simple 1:1 and monovalent 25:25 salt. In the latter cases, the {\em block} charge
      model has been adopted. With one exception, the total concentration of
      positive, as well as negative, charged spheres in the bulk is 2 M.
      The case of a pure 100 mM simple salt solution is shown as reference.
}
  \label{fig:25block}
\end{figure}

Let us revisit a process similar to that we considered for neutral polyampholyte chains.
We imagine a 2 M 1:1 salt solution in which monovalent {\em block} charge 25:25 clusters are formed, to a gradually
increasing extent. This then suggests a gradual drop of the remaining ``free'' ions, which together
form a 1:1 salt. The corresponding cDFT prediction of surface forces are
displayed in Figure \ref{fig:25block}.
As before, we find a dramatic impact by the
presence of {\em block} polyampholytes. While we would expect that clusters of
fully separated blocks of charge are unlikely, even at high ionic strength, the results
may be viewed as a useful limiting case, at the opposite extreme to that of alternating charges.
We would expect that alternating clusters should be preferred far from the charged surface, but
the field from highly charged surfaces is expected to generate polarized (block-like)
clusters in their vicinity. We expect that in a real scenario that surface forces will 
display a behaviour between these two extremes.

\section{Conclusions}
We have demonstrated that a commonly adopted simplification, where
ion cluster effects on surface forces are accounted for simply by the concomitant drop
of the dissociated (free) ion concentration, may turn out to be inaccurate.
This is because the ion clusters themselves contribute substantially
to the interactions.  Neutral clusters, modelled as chains of alternating 
charge (which may be prevalent due to the small Coulombic cost of creating them \cite{Hartel:23})
can have a significant effect due to their dielectric response.
Moreover, the massive dipolar response we have observed for block architectures 
suggests that perhaps even a modest anisotropy of the ion cluster charge distribution may
have a strong impact on the interaction between charged surfaces.

As already mentioned, our current models would benefit from
the incorporation of polydispersity, as well as the ability of the
chains to alter the connectivity in response to the presence of fields.
These effects will be accounted for in future work, utilising 
cDFT formulations that we have previously
developed for similar systems\cite{Woodward:08a}.

We have mainly discussed the polyampholytes as crude models of ion clusters, but it should
be noted that our findings have broader implications. Given a solution of relatively high
ionic strength, it would be quite feasible to {\em synthesize} polyampholytes
with a {\em block} charge structure, for instance using established biochemical approaches, and suitably chosen
peptides. Such {\em block} polyampholytes could then serve as stabilisers of colloidal dispersions, and
their impact might be modulated by appropriate adjustments of pH. Moreover, the
strong dipolar response that we observe is likely to be relevant
in solutions containing ``patchy'' colloidal particles, such as proteins. 

\begin{acknowledgments}
  J.F. acknowledges financial support by the Swedish Research Council.	
\end{acknowledgments}

\section*{AUTHOR DECLARATIONS}

\subsection*{Conflict of Interest}

The authors have no conflicts to disclose.

\subsection*{Author Contributions}
\textbf{David Ribar:} Conceptualization, Formal Analysis, Validation, Visualization,
Writing/Review \& Editing. \textbf{Clifford Woodward:} Conceptualization, Methodology,
Writing/Review \& Editing. \textbf{Jan Forsman:} Conceptualization, Formal Analysis,
Funding Acquisition, Methodology, Software, Supervision, Writing/Original Draft
Preparation, Writing/Review \& Editing. 

\section*{Data Availability Statement}
The data that support the findings of this study are available from the 
corresponding author upon reasonable request.

\appendix

\section{Branched polyampholyte structure}

\subsection{Branched neutral chains (alternating charge)}
\begin{figure}[!h]
	\centering
	\includegraphics[scale = 1]{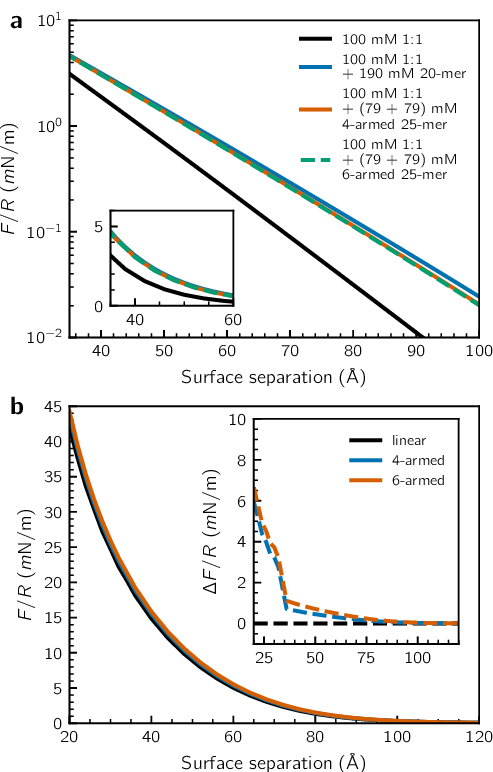}
	\caption{\textbf{a} Interactions between equally charged surfaces, in the presence of
					$X$ mM $r:r$ polyampholytes, and 100 mM simple 1:1 salt.
					For a given degree of polymerisation  $r$, $X$ is chosen in a way that is commensurate
					with clustering of a 2 M simple salt solution. However, a neutral branched chain also
					contains an uncharged central monomer. Moreover, the branched chains form a mixture
					of polymers with cation/anion end charges (although all branches are overall electroneutral).
					This means that (say) a 25-mer cluster contains 24 charges and that
					there are two different versions of such a neutral cluster. 
					\textbf{b} Surface forces, in the presence of 
	     			a mixture of 10 mM simple 1:1 salt and
					40 mM monovalent 25:25 salt (alternating charges). Apart from
					results for linear chains, we
					 provide results for branched (star-like) 25-mer polyampholyte models,
					 with 4 and 6 ``arms''. The insert present the difference from the linear surface forces, to
                                         highlight the structure effect.
%					\textbf{c} Surface forces, in the presence of pure 1:1 salt solutions (1 mM and 11 mM), and
%					in a mixture of simple 1:1 and monovalent 25:25 salt. In the latter case, the {\em alternating} charge  
                                        %					model has been adopted.}
                                        }
	\label{fig:A1}
\end{figure}

It turns out that, at least within the mean-field approximation, there
is no difference of practical importance between the multipole response from
linear or branched chain, at a similar degree of polymerisation. 
This is illustrated in Figure \ref{fig:A1}\textbf{a}.

This lends support to the hypothesis that a linear polymer model of ion clusters, which might 
appear unrealistically ``stretched'', actually is able to incorporate
the most salient physics, at least for models of neutral polyampholytes with alternating charges.
In other words, the overall multipole is quite similar (and significant), from models
using branched polymer architectures.

\subsection{Branched monovalent chains (alternating charge)}
Here we evaluate effects from making the charged monovalent
polymer models more ``compact'',and arguably more cluster-like, by
using branched (star-like) backbone descriptions. As we see in Figure \ref{fig:A1}\textbf{b},
the branched chains mediate stronger but more short-ranged surface forces. 
In the displayed case (40 mM monovalent 25-mers plus 10 mM 1:1 salt), the threshold
separation is about 70 {\AA}, below which the branched model produces a stronger repulsion.

%\subsection{Monovalent polyampholytes with alternating charge, plus a low concentration of simple salt}
%In Figure \ref{fig:A1}\textbf{c}, we note that the double-layer repulsions in dilute
%solutions of simple 1:1 salts, are considerably weaker, at short and intermediate
%separations, than those
%that also contain alternating charge clusters. This lends further support to
%the conclusion that the formation of ion clusters has broader implications
%to surface interactions than is implied simply by the accompanying drop of the ionic strength.

%On the other hand, on the insert of Figure \ref{fig:A1}\textbf{c} we demonstrate that,
%for an alternating charge structure of the polyampholytic chains, the decay
%at very long range agrees with predictions for simple salt solutions at the same overall ionic
%strength.

\footnotesize{}
\baselineskip=10pt
\bibliography{poly}
\end{document}